# The effect of network structure on innovation initiation process: an evolutionary dynamics approach

Afshin Jafari[1], S. Peyman Shariatpanahi[2], Mohammad Mahdi Zolfagharzadeh[1], Mehdi Mohammadi[1]

[1] Faculty of Management, University of Tehran, Jalal Al-e-Ahmad Ave., Nasr Bridge, Tehran, Iran, P.O. Box: 14155-6311.
[2] Institute of Biochemistry and Biophysics, University of Tehran, Enghelab Ave., Tehran, Iran P.O. Box: 13145-1384.

**Abstract.**
In this paper we have proposed a basic agent-based model based on evolutionary dynamics for investigating innovation initiation process. In our model we suppose each agent will represent a firm which is interacting with other firms through a given network structure. We consider a two-hit process for presenting a potentially successful innovation in this model and therefore at each time step each firm can be in on of three different stages which are respectively, Ordinary, Innovative, and Successful. We design different experiments in order to investigate how different interaction networks may affect the process of presenting a successful innovation to the market. In this experiments, we use five different network structures, i.e. Erdős and Rényi, Ring Lattice, Small World, Scale-Free and Distance-Based networks. According to the results of the simulations, for less frequent innovations like radical innovation, local structures are showing a better performance comparing to Scale-Free and Erdős and Rényi networks. Although as we move toward more frequent innovations, like incremental innovations, difference between network structures becomes less and non-local structures show relatively better performance.

## 1 Introduction

In today's world, innovation and suitable innovation strategy is crucial for firms to compete in their current market, to enter new markets and to have a sustainable competitive advantage [1, 2]. Many different definitions for innovation exists in the literature (see [3]), while we are not going to investigate them in detail, but it should be mentioned that innovations are not always significant advancements regarded as "radical" innovations; they may be "incremental" innovations which are relatively small but effective changes in technical knowledge of technology [3]. Modeling the process of presenting a successful innovation, as a social phenomenon, can be of great importance since it can help to better understanding of the phenomenon. Different methods and approaches like complex networks, evolutionary dynamics and agent-based modeling have already shown to be very helpful to model the process [4–6]. In this paper, we combined these methods to have a multidimensional view of the process of presenting a successful innovation.

In last decades, complex networks have shown a great potential for describing many different natural and social phenomena. These studies varies in different subjects, for example analyzing networks among students [7, 8], scientific collaborations [9], and innovation networks [4, 5, 10]. Analyzing network structures finds its most importance when we are dealing with social systems in which through different actors' interactions some macro behavior emerges; these emergent properties are direct results of the way actors are connected and interact. Network science consists of tools for investigating these structures and their effect on emergence of a particular behavior [11]. It should be mentioned that in literature innovation networks mostly refer to cooperation networks between different firms. However, having more general perspective, we can consider interaction networks between firms which can be relevant to cooperation, direct competition, imitation or any other relation that two firms are directly affected from each other's activities.

Evolutionary dynamics has shown to be a very strong tool to study social systems dynamics. There are three basic concepts in evolutionary dynamics: replication, selection, and mutation [12]. Reproductive population is a basis assumption in almost every evolutionary process. Natural selection can be regarded as differences of reproduction that maintains the fittest types while dismisses others. Mutation which is failure in replication process, is responsible for generating new types (variations) which will be evaluated in selection process [12]. For modeling evolution in finite population, Moran process presents a very simple solution [13]. In this process, suppose we have $N$ agents which are divided into more than one categories. At each step two random agents will be chosen, one of them will die and another one will reproduce itself. This newcomer offspring which has the same type as her mother, will take place of the dead agent and therefore total population will remain constant. The probability of selecting a random agent from a particular type, is proportional to that type's frequency and fitness[12, 13].

One of the most well-known evolutionary dynamical models in biology, which is also basis of our model, is two hit hypothesis for retinoblastoma which Knudson proposed in 1971 [14]. According to this hypothesis, Knudson argued that in order to one develop retinoblastoma, two genes, relevant to maternal and paternal chromosomes, in a cell must become defective and therefore it takes two mutations or two hits. These three states of cell are called respectively, wild-types, intermediate mutants, and double-mutants [15]. Three fitness parameters $f_1$, $f_2$, and $f_3$, are defined respectively for each state which can be interpreted as relative probabilities that a particular type will be chosen for reproduction. By comparing fitness of intermediate mutant and wild-type three different cases may occur as demonstrated in Fig. 1.

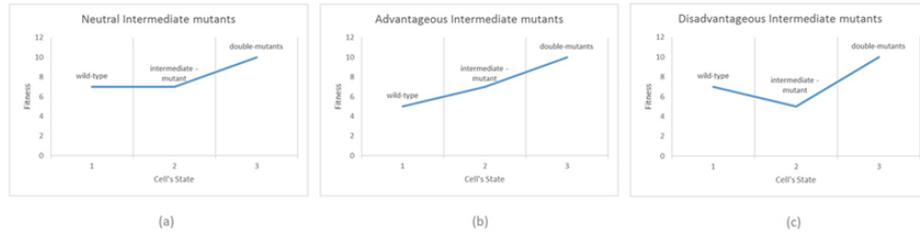

**Fig. 1.** Different types of mutation based on intermediate mutant in two-hit model. (a) neutral intermediate mutant, (b) advantageous intermediate mutant, and (c) disadvantageous intermediate mutant.

The same process can be considered for a firm with higher fitness in market if we assume that the way for a firm to be successful in a competition is through the path of innovation: Ordinary firms (wild type) become innovative (intermediate mutant) by investing on R&D section, although it may have a heavy cost; finally, this investment may bring them a huge profit (double mutants). This assumption is based on Porter's claim that in modern economy, because of highly connected world that we have today, things like being near to the source of inputs or having access to cheap labor are not effective competitive advantages anymore[16]. He argues that what derive today's economy is productivity and this requires continual innovation.

There are also numerous studies which demonstrate that there is a positive relation between R&D investments and firms innovation and growth [17, 18]. According to these studies and also models for innovation's life cycle like [19] which consider activities like market researches and prototyping as a crucial step before presenting an innovation, it seems reasonable to roughly divide the process of presenting a successful innovation into two steps, like what we did in our model which is presented in next section. It is also worth mentioning that in literature, different innovation strategies like product, process, market and organizational innovations are defined [3]. However, in this paper and in our basic model, we did not differ between different innovation strategies and consider all of them and their possible combinations as firms' decision to be innovative.

In this paper, we aim to study a special aspect of innovation process called "innovation initiation". There are several innovation diffusion studies, like [20], where researchers assume that there is a potentially successful innovation and investigate how that innovation will diffuse. There are also some models like Technology Acceptance Model [21] which suggest some features that an innovation must have to be successful. However, studying the process of presenting that successful innovation, which we call it innovation initiation, is a missing link in innovation studies. In this paper, we present an Agent-Based Model based on evolutionary dynamics, for investigating innovation initiation within different interaction network structures between firms. The main answered problem is how different interaction network structures may affect the process of innovation initiation. We apply Knusdson's two hit model for cancer to innovation initiation process. Finally, according to the model results, we propose optimal network structure considering both initiation and diffusion of innovation.

## 2   Method

In order to study the process of presenting a successful innovation in a network of competing firms, we choose agent-based modeling as our methodological approach. In our model, each agent represents a firm that is competing within a market. Each of these agents or firms can be in three different stages, Ordinary, Innovative, and Successful, which are congruous with three states of cells mentioned in section 1. If a firm's state is Ordinary, it means that this firm's strategy is to present stable and well established products or services with conventional methods or in other words, business as usual. These firms are playing in a safe and stable field with low risk of failure and low R&D costs. Despite these advantage, because of high level of competition in conventional markets the profit margin is low. Therefore, the fitness of this stage is expected to be neither high nor low.

Innovative state, represents a class of firms which are struggling to escape from that intense competing of previous state. In order to do that, these firms invest more in developing new and innovative products and services. These efforts toward innovativeness comes with a cost of accepting the high level of uncertainty. Therefore, this state is a transitional state from being an ordinary firm which is struggling to stay in a highly competitive market, toward a distinct and unique firm; the third state or Successful state as we defined in our model. This third and final state represents the situation that a particular firm was successful in developing and commercializing a new product or service. Firms in this state are enjoying high profit margin, high market share and even having some sort of monopoly. This is similar to what Kim and Mauborgne called "Blue Ocean" [22] or what Joseph Schumpeter called "Monopoly Profit" [23]. Thus, we can expect the fitness for this state to be considerably higher than other two states. We demonstrate these three states and possible transitions between them in Fig. 2.

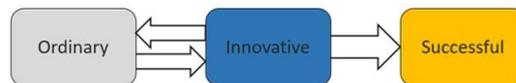

**Fig. 2.** Three different states for a firm in model and their succession

We suppose that spreading of strategies in population occurs with imitation process. This means that each agent must either decide to imitate strategy of someone else or keep its own strategy [24]. For finite number firms, $N$, we consider Moran process as our basic evolutionary process for selection. In order to model selection, we assume that, by a rate called "death rate" ($v$), each firm (Ordinary or Innovative) enters the situation in which it must rethink its strategy. At this point, firms must decide to be innovative and invest more money on R&D or to continue business as usual (see Fig. 2). According to the Moran process, the transition probability is dependent to different type's frequencies in that firm's network and their relative fitness. In our simulations we consider the neutral intermediate mutant (Fig. 1) case for simplicity and the fitness of innovative firms is assumed to be the same as ordinary ones. For instance, if we consider an Ordinary firm and denote the number of its connected firms in "innovative" state by $i$, and total number of connections of that firms by $N$, then the probability for that firm to decide to change its state from Ordinary to Innovative is equal to $i/N$. Mutation as another aspect of evolution, is considered in our model by a particular rate called "mutation Rate" or $u$. In the basic model which is presented here, for simplicity we assume that the only way for a firm to reach Successful state is by mutation from Innovative state to Successful state with predefined mutation rate. Also we did not consider back-mutations in our model.

In order to conform to reality, we need to explain what are our model's concepts instances and confirmations in real world. Death Rate indicates how fast firms can change their competing strategy or in other words how flexible firms are. Things like persistence of innovation activities, sunk costs, competition intensity, and time to develop an innovation are some factors which may affect this flexibility and subsequently Death Rate. Mutation Rate, depends on the level of innovation and also the type of industry or science. For example if we consider incremental innovation which happens almost continuously within different technological fields, this rate will be very high, but for radical innovations which happens infrequently, mutation rate would be much smaller [25]. Also, rate of technological change differs from sector to sector. Some sectors are characterized by rapid change and radical innovations, others by smaller, incremental changes [26].

We have executed simulations for $N = 2500$ agents and five different network structures with a wide range from completely random structures to restricted local structures. In the following, we describe these structures and how we apply them to our proposed model. Also in order to create and simulate the model, we have used AnyLogic software which allows us to create agent based models using Java programming language in a relatively user friendly environment.

### 2.1 Erdős and Rényi Random Network

One of the most well-known random structures is the model presented by Erdős and Rényi. In this network, we have a definite set of nodes, $N$, and each pair of nodes, independent of other connections, will be connected with a given probability of $p > 0$. One can easily show that for this process, degree distribution of any given node will follow binomial distribution and for large $N$ and small $p$, it can be approximated by Poisson distribution. In order to create this model in AnyLogic, we used its predefined Random network structure in which we have to specify the average number of connections for each node. In our simulations we assumed that this number is equal to 8 (Fig. 3(c)). From now on, for convenience and also in order to comply with the appellations used in AnyLogic, we will refer to structures based on Erdős and Rényi model, as Random networks.

### 2.2 Ring Lattice Network

This structure is called with different names like regular networks [27] or cliquish networks [28], but here, according to this network's name in AnyLogic, we refer to it as Ring Lattice network. In this structure, if we place agents on a ring, each one is connected to equal number of its neighbors on both sides (Fig. 3(a)). All links are local and there is no long range connection in this network structure. We assumed that each agent is connected to 8 other agents, 4 in each direction. We used AnyLogic pre-defined Ring Lattice network standard structure to create this network.

### 2.3 Small World Network

Watts and Strogatz suggests an algorithm to create this network [29]. According to this algorithm, first we have to create a Ring Lattice network. Then with a given probability, each link will be rewired to a randomly chosen agent. Therefore, this network is a combination of Ring Lattice and Random networks. Neighbor Link Probability is a probability which each link will remain local. For example if this probability is equal to 0.95, then with probability of $(1 - 0.95)$, each link would be rewired to a randomly chosen agent. We used the pre-defined Small World network in AnyLogic assuming 8 connections for each agent and 95 percent of the connections to be local, however we have changed this default percentage in some experiments.

### 2.4 Scale-Free Network

Many nodes with low degree and very few nodes with high degree are some of the main characteristic of this network. Degree distributions of nodes in this networks follows Power-Law distribution. One of the most well-known algorithms for generating Scale-Free networks is Barabási-Albert model [30]. This model is based on preferential attachment mechanism which means the preferability of a node is according to how many connections it has. Due to Barabási-Albert algorithm, first we will begin the network with $m_0$ initial nodes. At each steps, new nodes are added and get connected to $m < m_0$ existing nodes with a probability that is proportional to the number of links that the existing nodes already have. In our model, we have used the predefined Scale-Free network in AnyLogic in which, total population $N = 2500$, $m_0 = 2$ while parameter $m$ is calculated so that the average degree of the nodes would be 8. One can show for this network, total number of connections, $M$, would be equal to

$$M = 2 \times m \times (n - m) \tag{1}$$

Using Eq. 1, we find that $m = 4$, will approximately create a network with a total number of connections the same as Random network discussed before. Therefore we used $m = 4$ in our model for this network to have a comparable results (Fig. 3(d)).

### 2.5 Distance-Based Network

In Distance-Based networks, agents are located on a square lattice. Each agent will have 8 immediate neighbors which are connected to it with horizontal, vertical and diagonal links. We assume a periodic boundary condition as illustrated in Fig. 3(e).

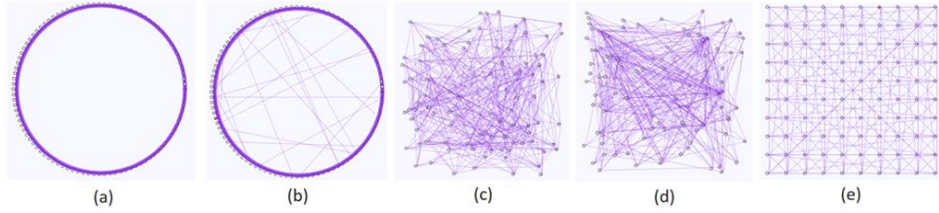

**Fig. 3.** Different network structures for $N = 100$ and average number of $connections = 8$. (a) Ring Lattice network, (b) Small World network with neighbor link probability $= 0.95$, (c) Random network, (d) Scale-Free network with $m = 4$, (e) Distance-Based network.

### 2.6 Simulations

Different simulations were performed in order to understand different aspect of model. In these experiments, we suppose that $N = 2500$, $f_3 = 7$, and $v = 0.2$. We have also calculated different networks' parameters in a way that average degree of each node be equal to 8. Also since in this model we are only interested in relative average times of second mutation in different networks, so we consider artificial time unit for our simulations and thus times reported in this section have no particular units and has no meaning except their ratios and proportions with respect to each other.

First series of experiments, is aimed to realize how changing the first states' fitness and its different situations comparing to second state, will affect the speed of presenting innovation. In order to do that, experiments with, $u = 10^{-4}$ and $f_2 = 5$ were performed in which we started from $f_1 = 5$ and after each 1000 runs $f_1$ was increased by 0.5 until it reached $f_1 = 7$. Therefore, we will start from neutral intermediate mutant case (Fig. 1(a)) and end up with a relatively highly disadvantageous intermediate mutant (Fig. 1(c)).

In the second series of experiments the effect of network structures on the innovation process was investigated. In these experiments, we consider the neutral, $f_1 = f_2 = 5$ and disadvantageous intermediate mutant case, $f_1 = 6$ and $f_2 = 5$. We neglect the advantageous intermediate mutant case since it seems unreasonable to consider more fitness for a transitional stage. We stop the simulation whenever second mutation occurred and record the time of second mutation. We run this experiment 50 times for each network structure and mutation rate for both cases.

The different network structures cover a wide range of structures from random to localized. To see the effect of localization in the network structure, in the third series of experiments, we investigate how changing the probability of having local links with neighbors will affect the simulation results for Small World network. In these experiments, we consider neutral intermediate mutant case with $u = 10^{-5}$ and we change neighbor link probability after each 50 runs.

## 3 Results

Simulation results for the first series of experiments shows that as we expected, by making it harder for agents to have first mutation, or in other words, by making it harder to invest in developing innovations, average time for presenting a successful innovation will increase (Fig. 4).

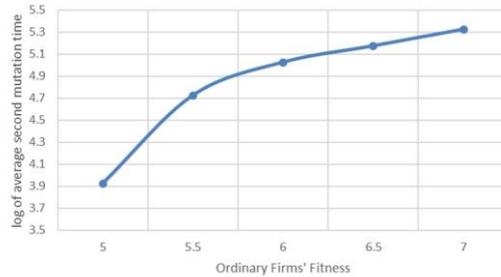

**Fig. 4.** Log of average second mutation time for different ordinary firms' fitness

Fig. 5 and Fig. 6, demonstrate the relative results for the second series of experiments. In these figures, for each value of mutation rate, we have demonstrated the relative average time for different networks to conduct the second mutation. We have also specified which network has the maximum average time in each particular mutation rate. By comparing these simulation times in both neutral and disadvantageous intermediate mutant cases, one can find how fast by increasing mutation rates, time to have second mutation will decrease.

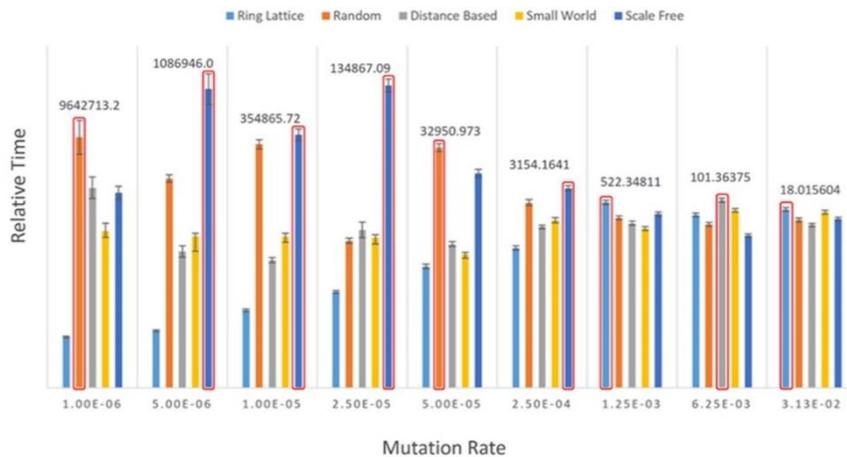

**Fig. 5.** Relative average time of second mutation for different mutation rates and different networks for neutral intermediate mutant, $f_1 = f_2 = 5$. Maximum average time for each mutation rate is mentioned on top of that section

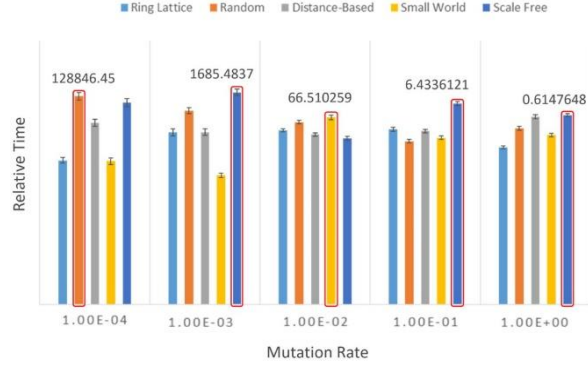

**Fig. 6.** Relative average time of second mutation for different mutation rates and different networks for disadvantageous intermediate mutant, $f_1 = 6$ and $f_2 = 5$.

The results of the third series of experiments is presented in Fig. 7. The figure shows that localizations decreases the needed time for innovation initiation.

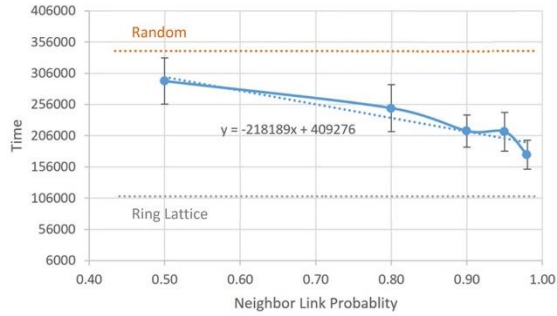

**Fig. 7.** Average second mutation time in Small World network for different neighbor link probabilities. $f_1 = f_2 = 5$ and $u = 10^{-5}$.

## 4 Discussion

The aim of the current research is to find the effect of network structure on the innovation initiation process and presenting a successful innovation to the market. Results indicate that this effect depends on the mutation rates. As we mentioned before, mutation rates depend on the type of innovation which we are dealing with. For example, for radical innovation this rate would be very low whereas for incremental innovations it is expected to be relatively high.

According to the results presented in Fig. 5 and Fig. 6, one can easily realize that for different mutation rates, or in other words for different innovation types, optimal network structure is different. For innovations with low frequency, e.g. radical innovations, which in our model we suppose they will have low mutation rates, local networks like Ring Lattice have shown a better performance. For these innovations, Random and Scale-Free networks in which connections are randomly distributed through the whole network, takes significantly more time to generate a successful innovation compared to Ring Lattice network. As it is demonstrated in the Fig. 5, by moving toward more frequent innovations, differences between networks reduce. However, interestingly, for these innovations it is local networks like Ring Lattice and Distance-Based network that show worst performances. In Fig. 6 we can see that there is a similar pattern to Fig. 5 in which at smaller mutation rates locals are significantly better and as we move toward higher mutation rates, this difference becomes negligible. Although for disadvantageous intermediate mutant case, even at highest possible mutation rate, still local structures are performing slightly better.

There are some models and theories in the literature of innovation which support the idea that the local connections result in a better performance in the sense of innovativeness. For example, concepts like geographically localized knowledge spillovers [31] or innovation clusters which is actually based on the assumptions that local connections increase firms innovativeness [1, 32]. These results should find more attention when we see in many innovations clusters, emerged naturally, there are limited hubs with high degree of centrality which is one of the signs for having Scale-Free network. For example, we can mention the role of Ericsson in Kista innovation clusters, Sweden [33]. Our results suggest that these structures suppress radical innovations.

It should also be mentioned that innovation initiation process is only one side of the coin which we discussed here. One of the main indicators that demonstrates the successfulness of an innovation, is its diffusion. Of course diffusion plays role in our model: when we consider an innovation to be successful, we actually mean that it has every requirement to diffuse within the network. However, since the simulation is stopped whenever a potentially successful innovation emerges, the diffusion process is not really investigated. By considering concepts like average shortest paths we can roughly argue about the diffusion process within different networks. Average shortest path indicates how easily two arbitrary agents can access to each other. This parameter is smaller for Random and Scale-Free networks. According to the literature, in network structures with small average shortest paths, diffusion of idea and knowledge is faster [27] although sometime a little bit localness may help creating initial critical mass for diffusion [28].

In order to have an innovative economy, both initiation and diffusion of innovation must be encouraged. According to our results, we find that for less frequent innovations, local networks perform better; for these radical innovations, initiation is the bottle neck and diffusion doesn't play a significant role. In the case of frequent innovations, diffusion becomes more significant and random networks are preferred.

## 5  Conclusion

In this article, we look into the process of presenting a successful innovation or as call it "innovation initiation". Our main question is to realize the effect of different network structure on the process of innovation initiation. In order to answer this question, we combine different concepts from complex networks and evolutionary dynamics in cancer and present an agent-based model. By conducting multiple simulations with our model, we realize that network structures effect varies for different innovation types. For less frequent innovations like radical innovations, local structures perform significantly better whereas, in frequent innovations like incremental innovations, it was Random and Scale-Free networks which perform better.

In many cases networks in real world are not the optimal, based on our findings. For example, where innovation clusters emerged around big firms, we can see different signs for Scale-Free networks, like limited highly connected hub or weak ties. The proposed model suggests that radical innovations are suppressed in these networks. Our findings can give some insights to policy makers to how form up new innovation clusters and how to modify existing clusters' network in order to have more innovative firms.